\documentclass[12pt,preprint]{aastex}
\usepackage{natbib}
\usepackage{amssymb}
\usepackage{lscape}
\usepackage{longtable}
\usepackage{url}
\usepackage{epsfig}

\shorttitle{Evidence for GW dominated emission in GRB 200219A}
\shortauthors{L\"{u} et al.}
\slugcomment{}

\begin{document}
\title{Evidence for gravitational-wave dominated emission in the central engine of short GRB 200219A}
\author{Hou-Jun L\"{u}\altaffilmark{1}, Yong Yuan\altaffilmark{1}, Lin Lan\altaffilmark{2},
Bin-Bin Zhang\altaffilmark{3}, Jin-Hang Zou\altaffilmark{4}, Zong-Kai Peng\altaffilmark{3}, Jun
Shen\altaffilmark{1}, Yun-Feng Liang\altaffilmark{1}, Xiang-Gao Wang\altaffilmark{1}, and En-Wei
Liang\altaffilmark{1}} \altaffiltext{1}{Guangxi Key Laboratory for Relativistic Astrophysics,
School of Physical Science and Technology, Guangxi University, Nanning 530004, China;
lhj@gxu.edu.edu} \altaffiltext{2}{Department of Astronomy, Beijing Normal University, Beijing,
China} \altaffiltext{3}{Key Laboratory of Modern Astronomy and Astrophysics (Nanjing University),
Ministry of Education, Nanjing 210093, China; bbzhang@nju.edu.cn} \altaffiltext{4}{Department of
Space Sciences and Technology, Hebei Normal University, Shijiazhuang 050024, China}

\begin{abstract}
GRB 200219A is a short gamma-ray burst (GRB) with an extended emission (EE) lasting $\sim 90$s. By
analyzing data observed with the {\em Swift}/BAT and {\em Fermi}/GBM, we find that a cutoff
power-law model can adequately fit the spectra of the initial short pulse with $\rm
E_{p}=1387^{+232}_{-134}$ keV. More interestingly, together with the EE component and early X-ray
data, it exhibits plateau emission smoothly connected with a $\sim t^{-1}$ segment and followed by
an extremely steep decay. The short GRB composed of those three segments is unique in the {\em
Swift} era and is very difficult to explain with the standard internal/external shock model of a
black hole central engine, but could be consistent with the prediction of a magnetar central engine
from the merger of an NS binary. We suggest that the plateau emission followed by a $\sim t^{-1}$
decay phase is powered by the spin-down of a millisecond magnetar, which loses its rotation energy
via GW quadrupole radiation. Then, the abrupt drop decay is caused by the magnetar collapsing into
a black hole before switching to EM-dominated emission. This is the first short GRB for which the
X-ray emission has such an intriguing feature powered by a magnetar via GW-dominated radiation. If
this is the case, one can estimate the physical parameters of a magnetar, the GW signal powered by
a magnetar and the merger-nova emission are also discussed.

\end{abstract}


\section{Introduction}
The progenitors of short gamma-ray bursts (GRBs) are thought to be from compact star mergers
\citep[for a review]{Kumar2015}, such as neutron star$-$neutron star mergers
\cite[NS$-$NS,][]{Paczynski1986,Eichler1989}, or neutron star$-$black hole mergers
\cite[NS$-$BH,][]{Paczynski1991}. Moreover, such coalescence systems are also the main targets as a
strong source of gravitational waves \cite[GWs,][for a review]{Berger2014}. The first directly
detection of GW event (GW 170817) associated with short GRB 170817A was achieved by Advanced LIGO
and Virgo \citep{Abbott2017a,Abbott2017b,Goldstein2017,Savchenko2017,Zhang2018a}, and it opened a
new window to study the properties for such a catastrophic crash \citep{Abbott2017a}.

The remnant of NS$-$BH mergers must be a BH surrounded by an accretion torus. NS$-$NS mergers, on
the other hand, may result in a BH \citep{Popham1999, Wheeler2000,Lei2009,Liu2017}, or a rapidly
spinning, strongly magnetized NS \cite[known as a millisecond
magnetar,][]{Usov1992,Thompson1994,Dai1998a,Dai1998b,Zhang2001,Metzger2011,Bucciantini2012,Lyu2015,Chen2017}.
From the theoretical point of view, depending on the nascent NS mass and poorly know equation of
state of NS, the magnetar can survive from milliseconds to several days
\citep{Lasky2014,Li2016,Gao2016,Lyu2018}. Based on the lifetime of the magnetar, three types of
magnetars can be classified, namely hypermassive NS, which is supported by differential rotation
with 10$-$100 ms lifetime before collapsing into a BH \citep{Rosswog2003} , supramassive NS, which
is supported by rigid rotation with survival from tens of seconds to thousands before collapsing to
a BH \citep{Rowlinson2010}, and stable NS with much longer lifetimes
\citep{Dai2006,Gao2006,Zhang2013,Kumar2015}.

From the observational point of view, the X-ray ``internal plateau'' (a nearly fairly constant
emission followed by a steep decay with a decay slope $\alpha
> 3$)\footnote{Throughout the paper we adopt the convention $F_\nu \propto t^{-\alpha}
\nu^{-\beta}$.} in some short GRBs may be a ``smoking gun'' for a supramassive magnetar as the
central engine \citep{Rowlinson2010,Rowlinson2013,Lyu2015,Lyu2017}. Moreover, the extended emission
(EE) of GRB prompt emission and shallow decay phase is naturally explained as the energy injection
of the magnetar in GRB central engine \citep{Dai1998a,Zhang2001,Troja2007,Lyons2010,Lyu2015}. The
magnetar can be spun down by losing its rotation energy via GW radiation or magnetic dipole (MD)
radiations
\citep{Dai1998a,Zhang2001,Corsi2009,Fan2013,Giacomazzo2013,Lasky2014,Metzger2014,Ravi2014,Lyu2014,Sowell2019}.
\cite{Lasky2016} presented more details to derive the luminosity of X-ray evolution with time as
$t^{-1}$ and $t^{-2}$ when the energy loss is dominated by GW and MD, respectively \cite[also
see][]{Zhang2001}. \cite{Lyu2018} found out several long GRBs whose X-ray emissions are consistent
with the above scenario. However, up to now, no directly observed evidence shows that the energy
loss of supramassive magnetar from double NS mergers is dominated by GW quadrupole emission.

GRB 200219A is a short-hard burst without redshift measurement. Based on the properties of its
multi-wavelength data presented here, the central engine of GRB 200219A seems to be a supramassive
magnetar which originates from the merger of an NS binary. We consider that the magnetar lost its
rotation energy via GW radiation before collapsing into BH. We present our data reduction from {\em
Swift} and {\em Fermi} observations in \S 2. In \S 3, we present the details of physical
interpretation with a magnetar central engine for GRB 200219A. The calculations of GW radiation and
possible merger-nova of magnetar are presented in \S 4 and \S5. Conclusions are drawn in \S6 with
some additional discussion. Throughout the paper, a concordance cosmology with parameters
$H_0=71~\rm km~s^{-1}~Mpc^{-1}$, $\Omega_M=0.30$, and $\Omega_{\Lambda}=0.70$ is adopted.

\section{Data reduction and analysis}

\subsection{{\em Swift} data reduction}
GRB 200219A triggered the Burst Alert Telescope (BAT) at 07:36:49 UT on 19 February 2020
\citep{Lien2020}. We downloaded the BAT data from the {\em Swift} website\footnote{$\rm
https://www.swift.ac.uk/archive/selectseq.php?source=obs\&tid=957271$}, and use the standard
HEASOFT tools (version 6.12) to process the BAT data, and running the late ``convert" command from
the HEASOFT software release to obtain the energy scale for the BAT events. The light curves in
different energy bands and spectra are extracted by running batbinevt \citep{Sakamoto2008}. Then,
we calculate the cumulative distribution of the source counts using the arrival time of a fraction
between 5\% and 95\% of the total counts to define $T_{\rm 90}$. The time bin size is fixed to 128
ms in this case due to the short duration. The light curve shows a short-pulse with duration
$T_{\rm 90}\sim 0.48$ s (see Figure \ref{fig:BATGBM}). The background is extracted using two
time-intervals, one before and one after the burst. We model the background as Poisson noise, which
is the standard background model for prompt emission in BAT events. We invoked Xspec to fit the
spectra. For technical details, please refer to \cite{Sakamoto2008}. The time-averaged spectrum of
short-pulse is best fit by a simple power-law model with an index $0.76\pm 0.08$. Moreover,
\cite{Laha2020} report that the initial short-pulse seems to be followed by a soft EE component
lasting $\sim 90$ s, and maybe some even weaker emission until $\sim 300$ s. The X-ray Telescope
(XRT) began observing the field 67 seconds after the BAT trigger \citep{Lien2020}. We made use of
the public data from the {\em Swift} archive \footnote{$\rm
https://www.swift.ac.uk/xrt\_curves/00957271$}\citep{Evans2009}. The Ultra-Violet Optical Telescope
\cite[UVOT,][]{Roming2005} observed the field at $T_0+74$ s, but no optical afterglow was
consistent with the XRT position \citep{Siegel2020}.

\subsection{{\em Fermi} data reduction}
\begin{figure}
\centering
\includegraphics [angle=0,scale=0.4] {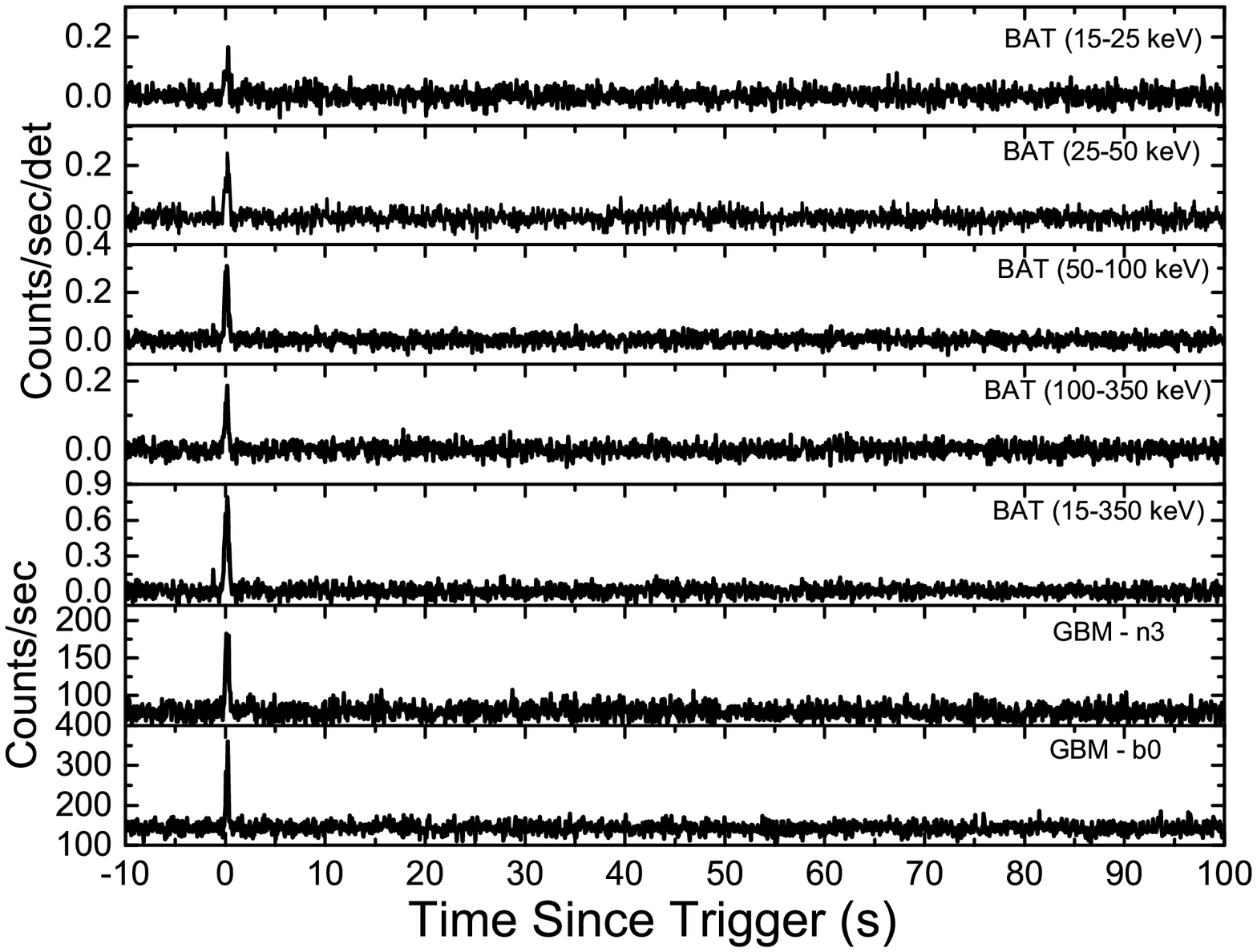}
\includegraphics [angle=0,scale=0.25] {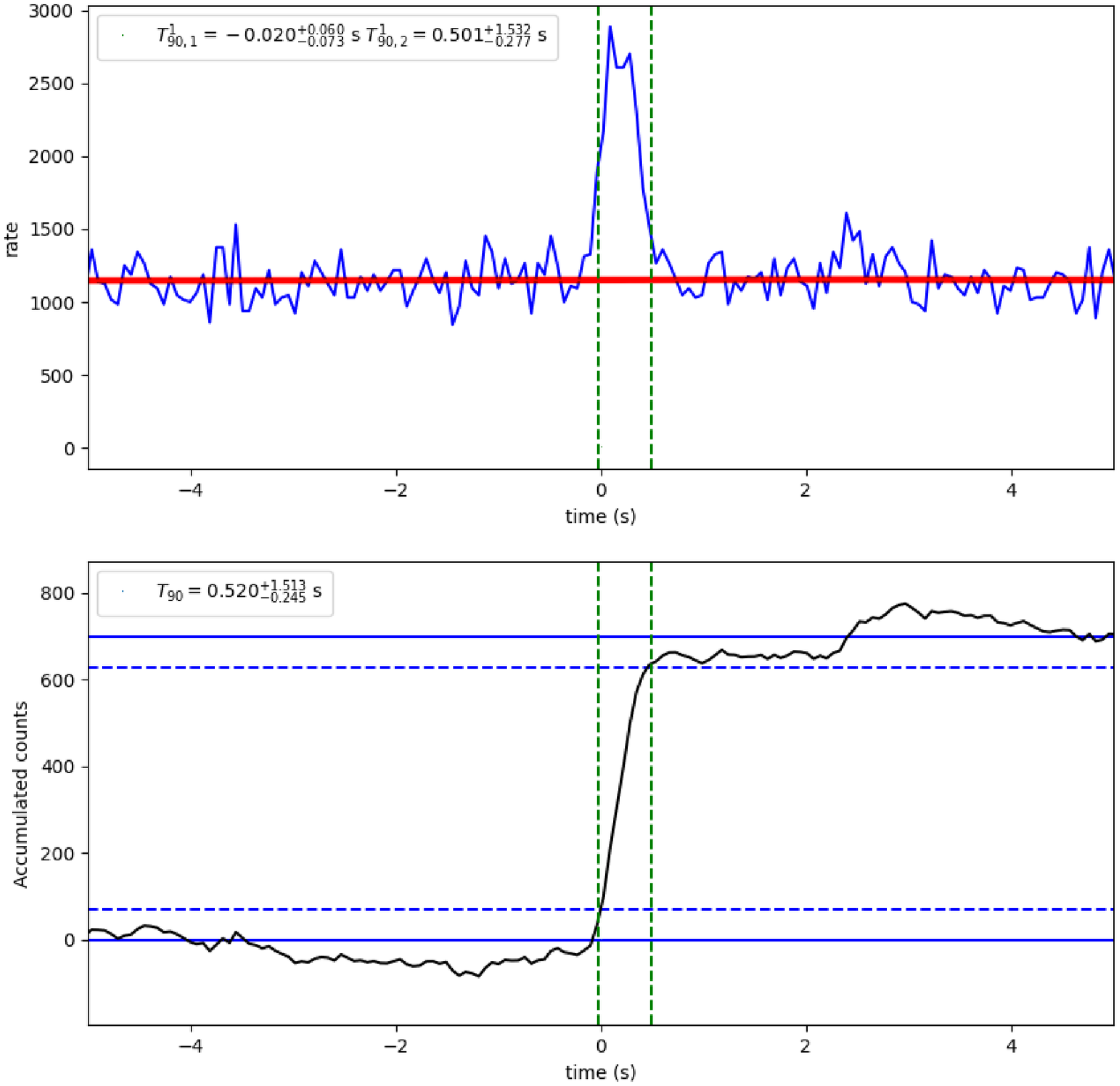}
\caption{{\em Left:}{\em Swift}/BAT and {\em Fermi}/GBM light curves of GRB 200219A in different
energy bands with a 128 ms time bin. {\em Right:} The determination of its $T_{90}$ for {\em
Fermi}/GBM. Green horizontal dashed lines are shown 5\% and 95\% of accumulated counts. Vertical dotted
lines
are drawn at the times corresponding to accumulated counts, which are used to define the $T_{90}$
intervals.}
\label{fig:BATGBM}
\end{figure}

The Fermi Gamma-ray Burst Monitor (GBM) was triggered and located GRB 200219A at 07:36:49.10 UT on
19 February 2020 \citep{Bissaldi2020}. GBM has 12 sodium iodide (NaI) and two bismuth germanate
(BGO) scintillation detectors are covering the energy range from 8 keV to 40 MeV
\citep{Meegan2009}. We downloaded the corresponding Time-Tagged-Event data from the public data
site of {\em Fermi}/GBM\footnote{{\bf $\rm
https://heasarc.gsfc.nasa.gov/FTP/fermi/data/gbm/daily/$}}. For more details of data reduction of
light curve and spectra procedure refer to \citep{Zhang2016}. The light curves of n3 and b0
detectors are shown in Figure \ref{fig:BATGBM}, it consists of a single emission episode. We
estimate $T_{90}$ of the burst according to the cumulative net count rate, and the duration of 90\%
total net counts is $T_{90}\sim 0.52$ s in 50$ - $300 keV with starting and ending time
$T_{90,1}\sim -0.02$ s and $T_{90,2}\sim 0.5$ (Figure \ref{fig:BATGBM}). The EE component of the
burst is not significant in the GBM temporal analysis.

\begin{figure}
\centering
\includegraphics [angle=0,scale=0.5] {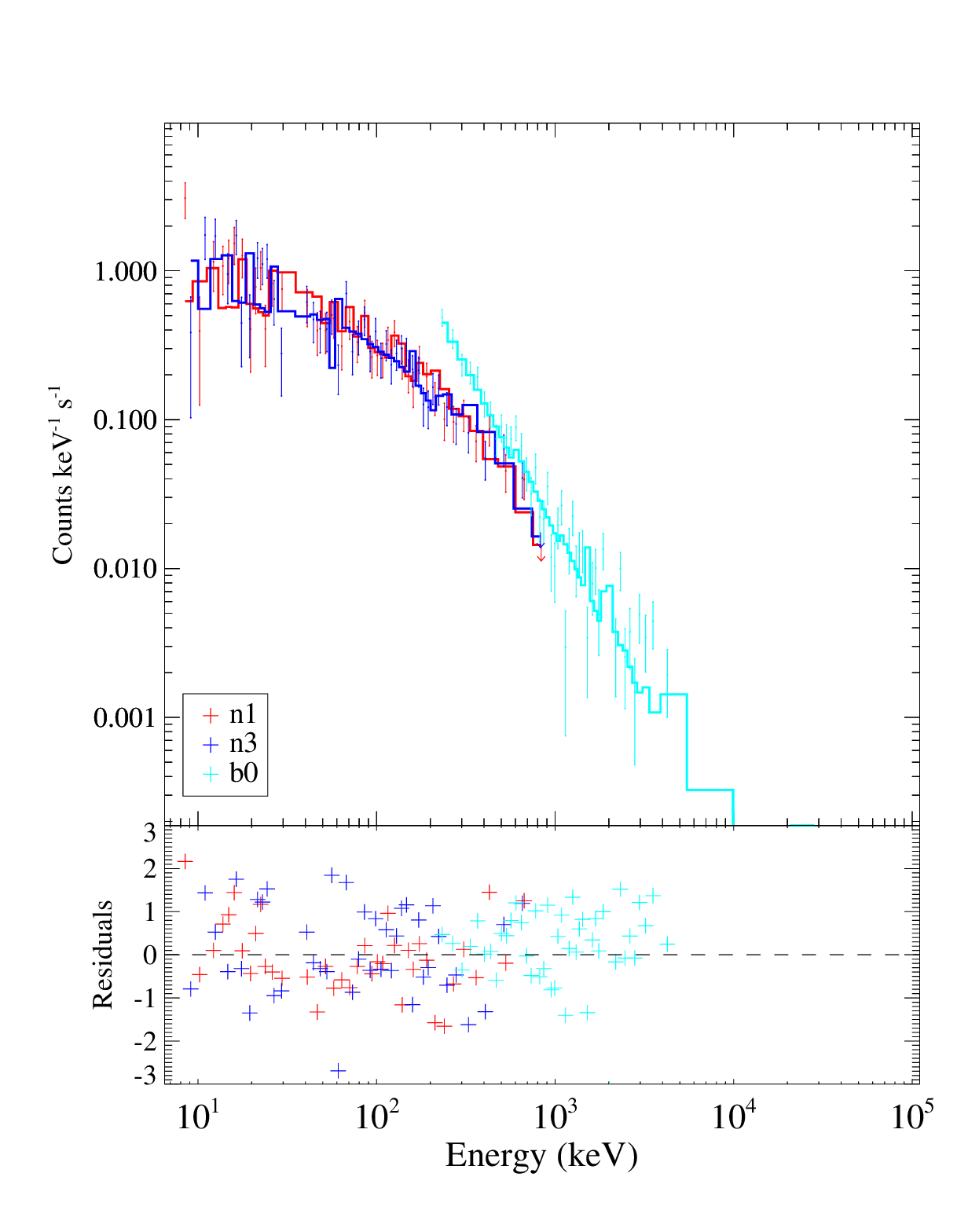}
\includegraphics [angle=0,scale=0.5] {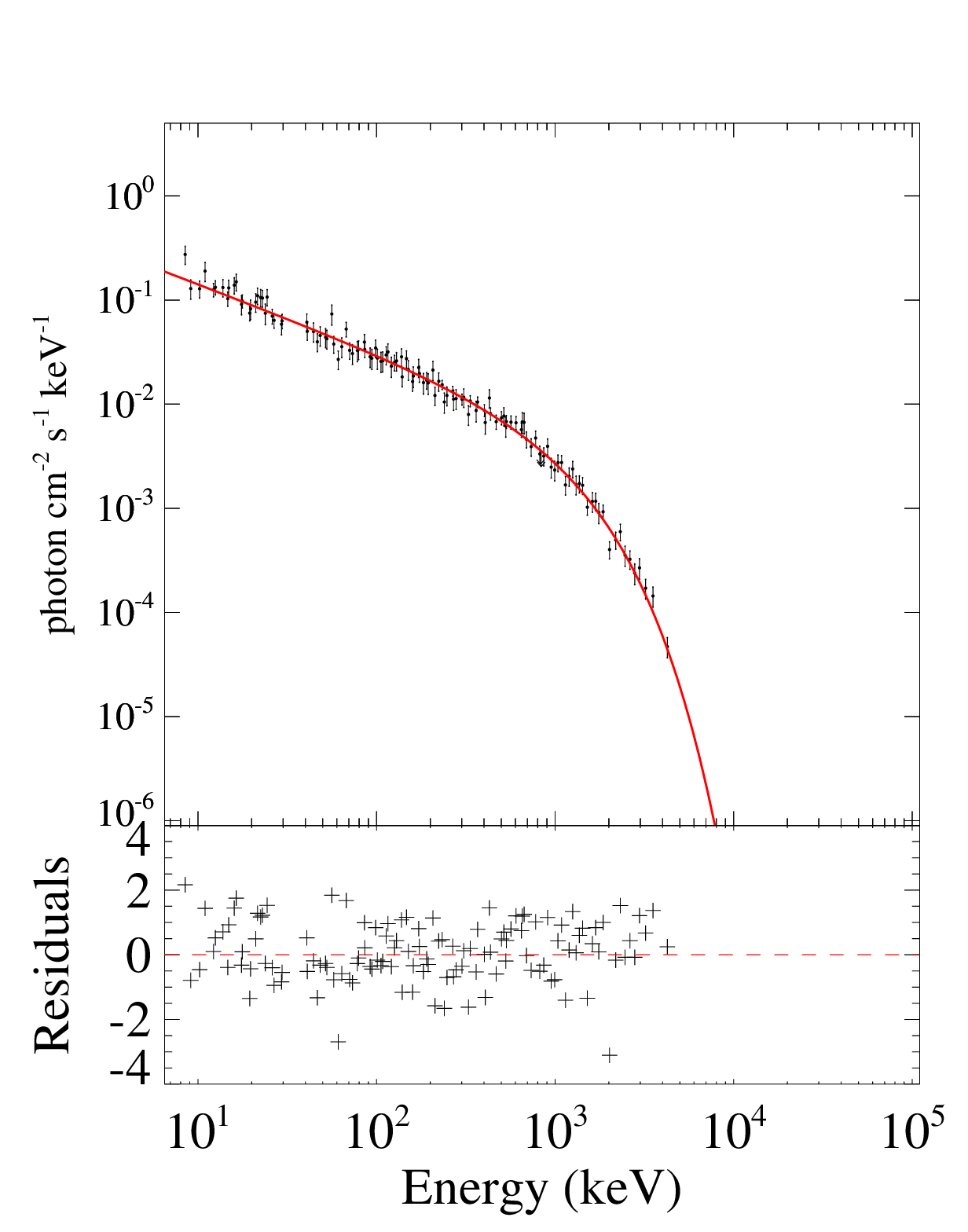}
\includegraphics [angle=0,scale=0.3] {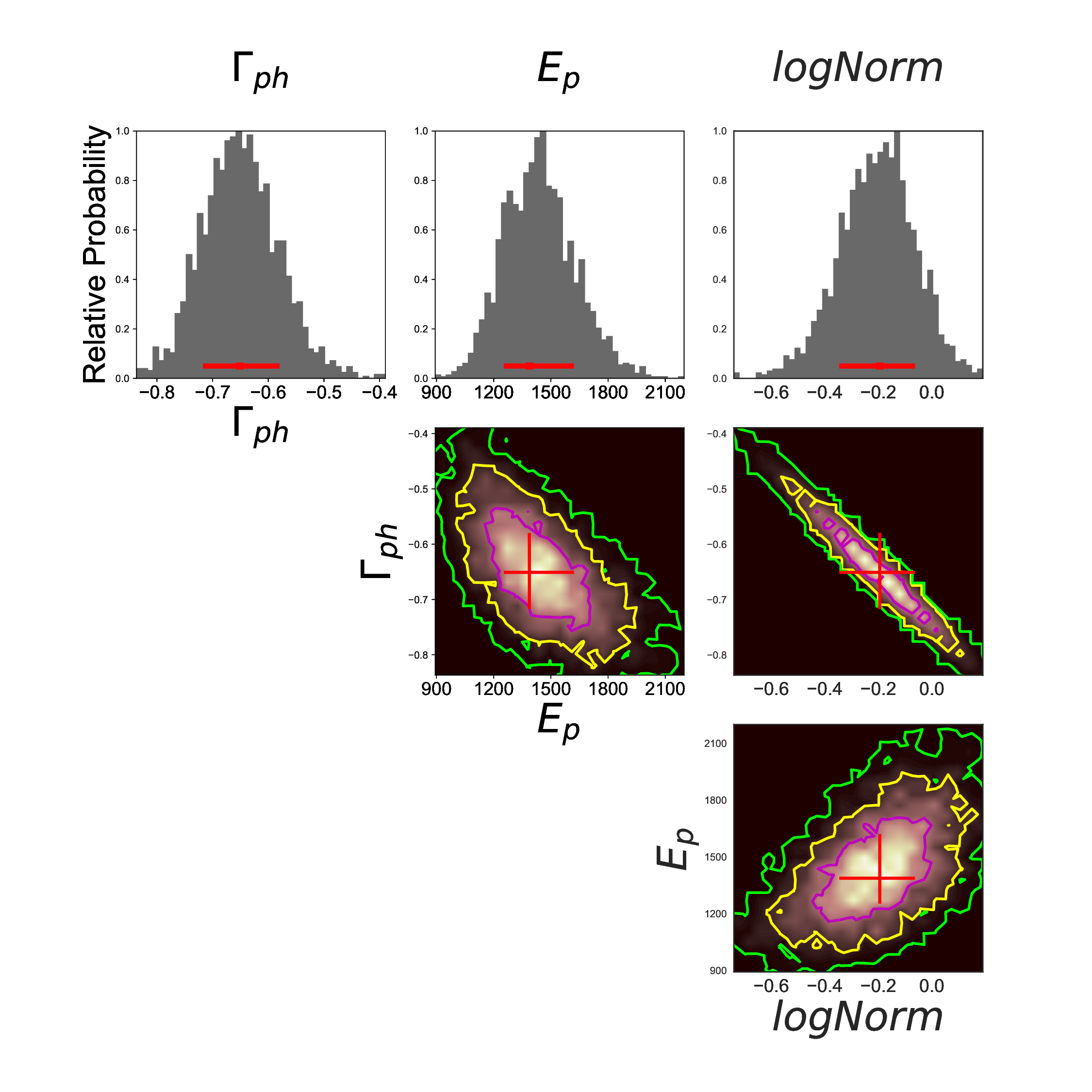}
\caption{The spectral fits of GRB 200219A with cutoff power-law model for
Fermi/GBM $T_{90}$. The count spectrum ({\em Top left}), photon spectrum ({\em Top right}),
as well as parameter constraints of the CPL fit. Histograms
and contours in the corner plots show the likelihood map of constrained parameters by using
our McSpecFit package. Red crosses are the best-fitting values, and pink, yellow, and green
circles are the 1$\sigma$, 2$\sigma$, and 3$\sigma$ uncertainties, respectively.}
\label{fig:SpecGBM}
\end{figure}

We also extract the time-averaged spectrum of GRB 200219A between $T_{90,1}$ and $T_{90,2}$. The
background spectra are extracted from the time intervals before and after the prompt emission phase
and modeled with an empirical function \citep{Zhang2011}, and the spectral fitting is performed by
using our automatic code ``{\em McSpecfit}'' in \cite{Zhang2018b}. Several spectral models can be
selected to test the spectral fitting of burst, such as power-law (PL), cutoff power-law (CPL),
Band function (Band), Blackbody (BB), as well as combinations of any two models. Then, we compare
the goodness of the fits and find that the CPL model is the best one to adequately describe the
observed data by invoking the Bayesian Information Criteria (BIC)\footnote{The BIC is a criterion
for model selection among a finite set of models. The model with the lowest BIC is preferred. The
BIC values of different model are presented as: 477(Band), 358(BB), 293(CPL), 505(PL),
369(Band$+$BB), 517(BB$+$PL), 304(BB$+$CPL).}. The CPL model fit is shown in Figure
\ref{fig:SpecGBM} for the count spectrum and photon spectrum, as well as parameter constraints of
the fit. It gives peak energy $\rm E_{p}=1387^{+232}_{-134}$ keV, and a lower energy spectral index
of $\Gamma_{\rm ph}=-0.65\pm0.07$. The best-fit parameters of CPL fits are listed in Table 1. The
estimated event fluence (1$ - 10^4$ keV) in this time interval is $5.08^{+1.27}_{-1.02}\times
10^{-6}~\rm erg ~cm^{-2}$.

\subsection{Statistical comparison of Burst and X-ray light-curve fits}
\begin{figure}
\centering
\includegraphics [angle=0,scale=0.45] {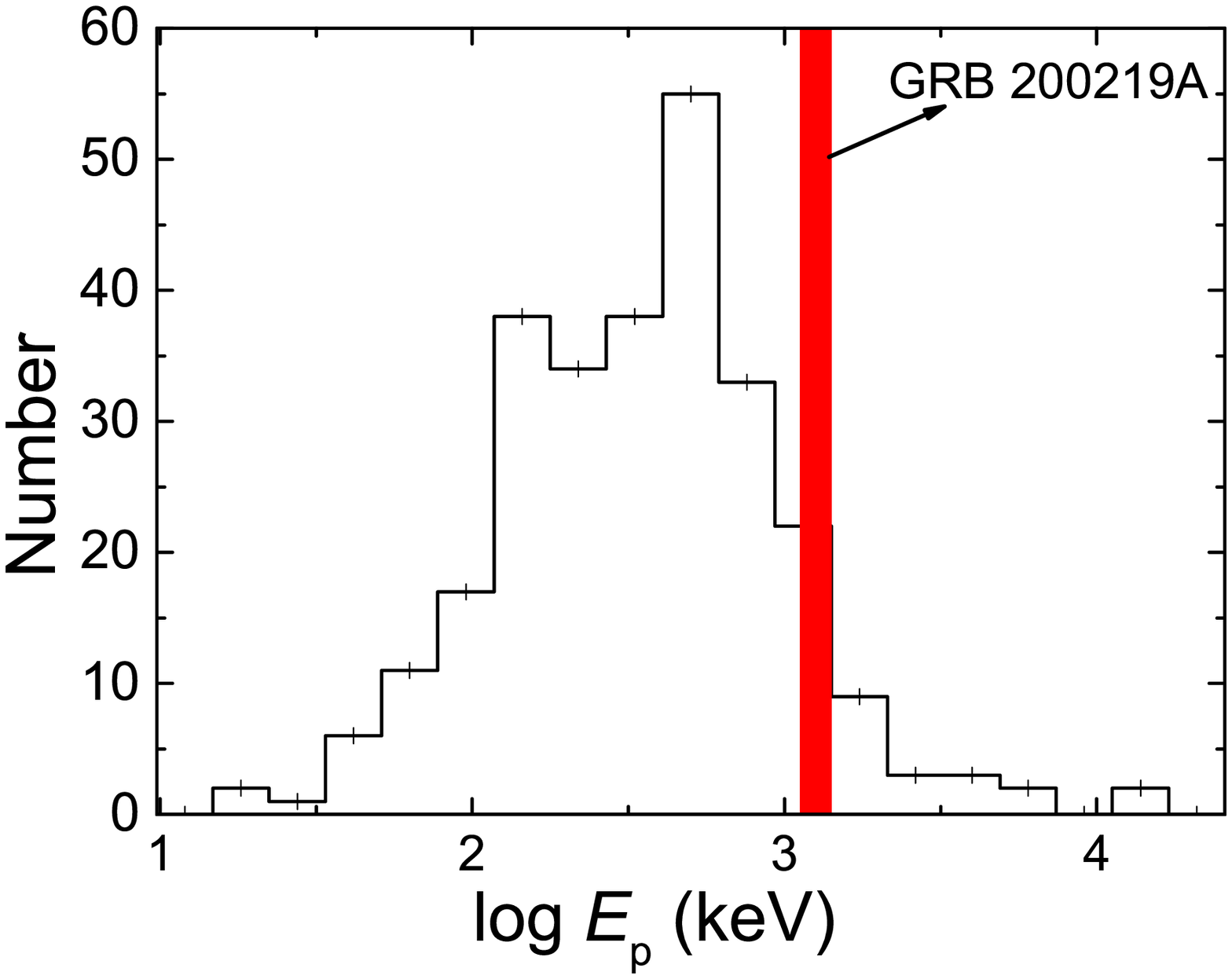}
\includegraphics [angle=0,scale=0.45] {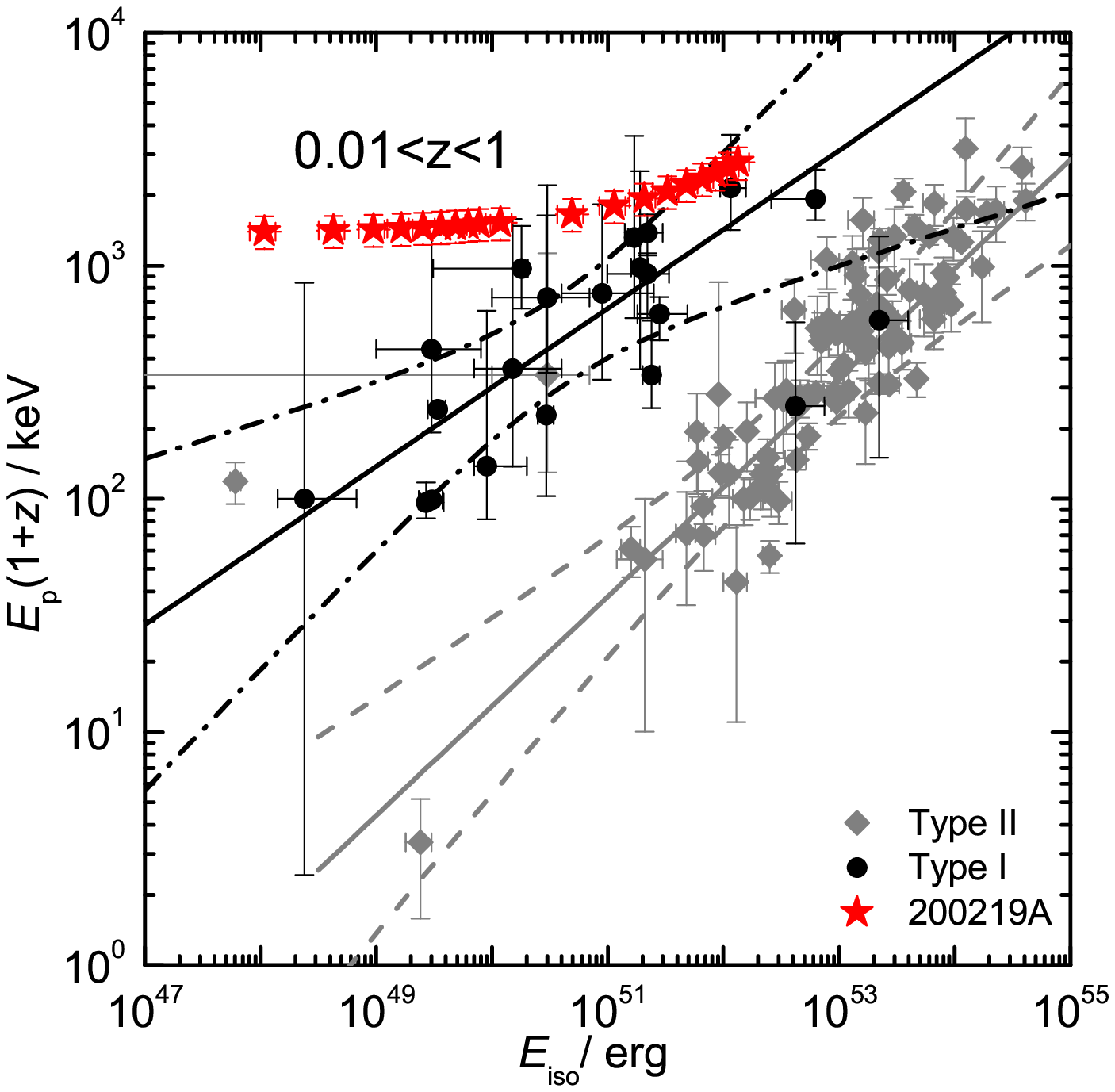}
\caption{{\em Left}: $E_{\rm p}$ distribution for GRB 200219A and other short GRBs observed by {\em
Fermi}/GBM. The $E_{\rm p}$ values of other short GRBs are taken from Lu et al. (2017). {\em Right}:
$E_{\rm p}$ and $E_{\rm iso}$ correlation diagram. Black points and grey diamonds are corresponding to
Type I and Type II GRBs, which are taken from Zhang et al. (2009). The red stars are the GRB 200219A with
pseudo redshift from 0.01 to 1. The redshift step is 0.01 from $z=0.01$ to $0.1$, and with step 0.1 from
$z=0.1$ to $1.0$. The best-fit $E_{\rm p}-E_{\rm iso}$ correlations for both
Type II (grey diamonds) and Type I (black points) GRBs are plotted (solid lines) with the
$3\sigma$ boundary (dashed line) marked.}
\label{fig:Comp}
\end{figure}

The observed properties of prompt emission for short GRB 200219A has a high peak energy and EE
component. By comparing the $\rm E_{p}$ value of GRB 200219A with that of other short GRBs observed
by {\em Fermi}/GBM, we find that the $\rm E_{p}$ of GRB 200219A is larger than that of most other
short GRBs, but still falls in the typical range (see Figure \ref{fig:Comp}). Moreover, no optical
counterpart associated with GRB 200219A means that the redshift is unknown. In order to check
whether GRB 200219A is an unusual event or not, we overplot GRB 200219A in the $E_{\rm p}-E_{\rm
iso}$ diagram \citep{Amati2002,Zhang2009} with pseudo redshift from $z=0.01$ to $z=1$ in Figure
\ref{fig:Comp}. The data of Type I and Type II GRBs\footnote{\cite{Zhang2009} proposed that Type I
and Type II GRBs are originated from compact stars merger and massive star core-collapse,
respectively. }, as well as fits are taken from \cite{Zhang2009}. We find that it an outlier of the
short GRB population for $z<0.5$, but is located well within the $1\sigma$ region for $z>0.5$.

The X-ray light curve of GRB 200219A seems to be interesting, it {\bf is} composed of several power
law segments. Firstly, we extrapolate the BAT (15$-$150 keV) data to the XRT band (0.3$-$10 keV) by
assuming a single power-law spectrum \citep{O'Brien2006,Liang2008,Evans2009,Li2012}. Then, we
perform a empirical fit to the light curve with a smoothed triple power law model,
\begin{equation}
\label{STPL}
F=(F_1^{-\omega_2}+F_2^{-\omega_2})^{-1/\omega_2},
\end{equation}
where $F_1$ and $F_2$ can be expressed as
\begin{equation}
\label{SBPL}
F_1=F_{0}\left[\left(\frac{t}{t_{\rm
b,1}}\right)^{\omega_1\alpha_1}+\left(\frac{t}{t_{\rm
b,1}}\right)^{\omega_1\alpha_2}\right]^{-1/\omega_1},
\end{equation}
\begin{equation}\label{PL}
F_2=F_1(t_{b,2})\left(\frac{t}{t_{b,2}}\right)^{-\alpha_3},
\end{equation}
where $t_{b,1}$ and $t_{b,2}$ are the two break times, $\alpha_1$, $\alpha_2$, and $\alpha_3$ are
the decay slopes before and after $t_{b,1}$, and after $t_{b,2}$, respectively. $\omega_1$ and
$\omega_2$ describe the sharpness of the break at $t_{b,1}$ and $t_{b,2}$, and
$\omega_1=\omega_2=3$ is fixed in our fits. The light curve fitting is shown in Figure
\ref{fig:XRT}, and fitting results are presented below, $t_{b,1}=(57\pm18)$ s, $t_{b,2}=(190\pm27)$
s, $\alpha_1=0.05\pm0.08$, $\alpha_2=1.18\pm0.15$, $\alpha_3=4.67\pm0.24$, $F_0=(5.08\pm
1.16)\times 10^{-9}~\rm erg ~cm^{-2}~s^{-1}$, and $\chi^2/dof=72/54$. One basic question is that
whether one power-law, or smooth broken power-law function can fit the the light curve well enough.
In order to test that, we also adopted one power-law function to fit the data, and find that the
$\chi^2/dof=75/32$, and even for a smooth broken power-law function with $\chi^2/dof=74/45$. So
that, we adopted a smoothed triple power law model to fit the data and infer the physical
parameters with the fitting results.

\section{Physical interpretation with magnetar central engine}
Magnetar as central engine of short GRBs are extensively discussed \citep{Dai2006, Gao2006,
Rowlinson2010, Metzger2011, Rowlinson2013, Kumar2015, Lyu2015}. Considering a newly born magnetar,
it is spun down via a combination of electromagnetic (EM) dipole and gravitational wave (GW)
quadrupole emission \citep{Shapiro1983,Zhang2001},
\begin{eqnarray}
-\frac{dE_{\rm rot}}{dt} = -I\Omega \dot{\Omega} &=& L_{\rm EM} + L_{\rm GW} \nonumber \\
&=& \frac{B^2_{\rm p}R^{6}\Omega^{4}}{6c^{3}}+\frac{32GI^{2}\epsilon^{2}\Omega^{6}}{5c^{5}},
\label{Spindown}
\end{eqnarray}
where $E_{\rm rot}=\frac{1}{2} I \Omega^{2}$ is the rotation energy of magnetar, $I$ is the moment
of inertia, $\Omega$, $P_0$, $B_p$, $\epsilon$, $R$, and $M$ are the angular frequency, rotating
period, surface magnetic field, ellipticity, radius, and mass of the neutron star, respectively.
The convention $Q = 10^x Q_x$ is adopted in cgs units.

\cite{Lasky2016} and \cite{Lyu2018} derived the electromagnetic luminosity as function of time for
both EM dipole and GW emission dominated energy loss, and the evolution behaviors are following,
\begin{eqnarray}
L_{\rm EM}(t) &=& L_{\rm em,0}(1+\frac{t}{\tau_{\rm c,em}})^{-2}, ~\rm (EM~dominated)
\label{Luminosity_EM}
\end{eqnarray}
\begin{eqnarray}
L_{\rm EM}(t) &=& L_{\rm em,0}(1+\frac{t}{\tau_{\rm c,gw}})^{-1}, ~\rm (GW~dominated)
\label{Luminosity_GWEM}
\end{eqnarray}
where $L_{\rm em,0}$ is the initial kinetic luminosity,
\begin{eqnarray}
L_{\rm em,0}&=&1.0 \times 10^{49}~{\rm erg~s^{-1}} (B_{p,15}^2 P_{0,-3}^{-4} R_6^6),
\label{spinlu_em}
\end{eqnarray}
$\tau_{\rm c,em}$ and $\tau_{\rm c,gw}$ are the characteristic spin-down time scale for EM and GW
dominated, respectively,
\begin{eqnarray}
\tau_{\rm c,em}\sim 2.05 \times 10^3~{\rm s}~ (I_{45} B_{p,15}^{-2} P_{0,-3}^2 R_6^{-6}),
\label{spintau_em}
\end{eqnarray}
\begin{eqnarray}
\tau_{\rm c,gw}\sim 9.1 \times 10^3~{\rm s}~ (I^{-1}_{45}\epsilon_{-3}^{-2} P_{0,-3}^4 ).
\label{spintau_gw}
\end{eqnarray}
Moreover, one can obtain the transition time ($\tau_{\ast}$) which point is from GW dominated to EM
dominated emission \citep{Zhang2001,Lasky2016,Lyu2018},
\begin{eqnarray}
\tau_{\ast}=\frac{\tau_{\rm c,em}}{\tau_{\rm c,gw}}(\tau_{\rm c,em}-2\tau_{\rm c,gw})
\label{transition_time}
\end{eqnarray}

The formation and evolution of magnetar central engine roughly can be described as following. A
possible remnant of double NSs merger is supramassive NS that is supported by rigid ration if the
equation of state of NS is stiff enough. The magnetar is going on losing its rotation energy via MD
or GW radiations to result in the magnetar spin-down due to its strong surface magnetic filed
and/or asymmetry of mass. If the energy loss of NS is initially dominated by MD radiation, the
luminosity evolves as $L\propto \sim t^{-2}$, and it survives until is collapsed into black hole
with a more steeper decay. Alternative, if the energy loss of NS is initially dominated by GW
radiation, the luminosity evolves as $L\propto \sim t^{-1}$ until MD dominated with $\sim t^{-2}$,
or with steeper decay when it collapse into black hole before switching the MD radiation.

Motivated by the above derivations, we find that the X-ray evolution behavior of GRB 200219A is
consistent with the magnetar central engine. The physical process can be described as follows: the
progenitor of short GRB 200219A originated from double NSs merger, a supramassive NS is produced
after the merger (if the masses distribution of the two NSs are perfection, or its equation of
state is stiff enough), accretion of the torus material into the NS may launch a jet, which powers
a short-duration GRB (the prompt emission of short GRB). The observed EE component or plateau
emission is from the energy injection of magnetar dipole radiation before the NS spin-down. After
tens of seconds, the newborn NS is spun down by losing its rotation energy via GW emission (the
observed $\sim t^{-1}$ segment), and then survive for hundreds of seconds before collapsing to a BH
(even steeper segment), but is not enough time to switch into EM dominated phase. Moreover, a
normal decay segment with slope $\sim 1$ at the later time is consistent with external shock model,
this component is the afterglow emission from the jet.

If this is the case, we find that the GRB 200219A possibly presents the first indirect evidence to
show the GW-dominated emission of supramassive NS in the central engine, this can be confirmed by
systematically searching for all short GRBs observed by {\em Swift}/BAT. However, with no measured
redshift it is difficult to evaluate the properties of supramassive magnetar; there pseudo redshift
values are adopted in our calculations. One is $z=0.01$ (corresponding to $\sim$40 Mpc luminosity
distance), that of GW170817/GRB 170817A \citep{Abbott2017a}. Another one is $z=0.1$ (corresponding
to $\sim$450 Mpc luminosity distance), which is close to the upper limit of the GW signal detected
by LIGO/Vergo \citep{Abbott2017b}. The third one is $z=0.5$ the central value of redshift
distribution for all short GRBs with z measurements \citep{Lyu2015}. Based on the derivations of
above, one has $\tau_{\rm c,gw}\simeq t_{b,1}/(1+z)$, and $\tau_{\ast}>t_{b,2}/(1+z)$. Together
with Eq. (\ref{transition_time}) and standard error propagation, one can roughly estimate
$\tau_{\rm c,em}>135/(1+z)$ s. On the other hand, by assuming the efficiency $\eta=0.1$, $\eta
L_{\rm em,0}\simeq 4\pi D_L^2 F_0$ ($D_L$ is the luminosity distance) at redshift $z=$0.01, 0.1,
and 0.5, one can estimate the upper limit of $B_{\rm p}$ and $P_0$ with equation of state GM1
\citep{Lasky2014,Ravi2014,Lyu2015}. The results are presented in Table 2. The comparison of the
magnetar parameters with other short GRBs is shown in Figure \ref{fig:XRT}. The derived magnetar
parameters of other short GRBs are taken from \citep{Lyu2015}, they invoked the observed X-ray
internal plateau of short GRBs to constrain magnetar parameters by assuming the energy loss from
dipole radiation.

\section{GW radiation of magnetar}
\begin{figure}
\centering
\includegraphics [angle=0,scale=0.6] {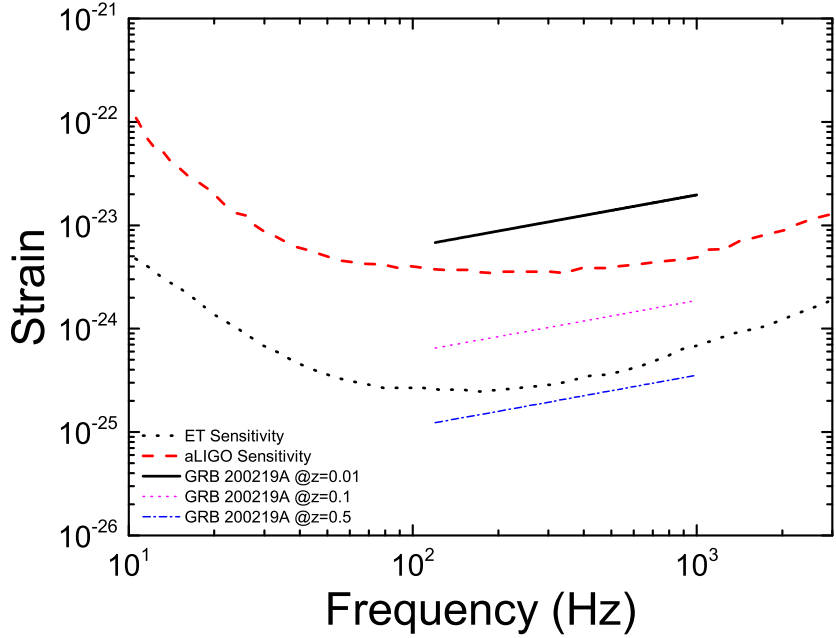}
\caption{GW strain evolution with frequency for GRB 200219A at pseudo redshift $z=0.01$ (black solid line),
$z=0.1$ (pink dotted line), and $z=0.5$ (blue dash¨Cdotted line). The black dotted line and red dashed line
are the sensitivity limits for aLIGO and ET, respectively.}
\label{fig:LIGO}
\end{figure}

If the energy loss of magnetar is dominated by GW radiation, one potential question is that how
strong is the GW signal of magnetar. Based on the Eq.(\ref{Spindown}), one can derive the
$\Omega(t)$ evolution as function of time \citep{Lyu2018},
\begin{eqnarray}
\Omega(t) = \Omega_{0}(1+\frac{t}{\tau_{\rm c,gw}})^{-1/4},
\label{Omega_GW}
\end{eqnarray}
and hence the GW frequency
\begin{eqnarray}
f(t) = f_{0}(1+\frac{t}{\tau_{\rm c,gw}})^{-1/4},
\label{fre}
\end{eqnarray}
where $f_{0}$ is initial GW frequency. So that, the amplitude of the GW signal decreases with time
as \citep{Fan2013,Lasky2016,Lyu2017}
\begin{eqnarray}
h_{\rm c} = \frac{1}{D_{\rm L}}\sqrt{\frac{5GIf_{0}}{2c^{3}}}(1+\frac{t}{\tau_{\rm c,gw}})^{-1/4}
\label{GWsignal}
\end{eqnarray}
Here, the $h_{\rm c}$ is characteristic gravitational wave amplitude. The GW signal of new born
magnetar at $t\sim 0$ is the strongest, so that the Eq.(\ref{GWsignal}) can be approximate to
following,
\begin{eqnarray}
h_{\rm c}&\approx& 8.22\times 10^{-24} \biggl(\frac{I}{10^{45}~\rm g\,cm^{2}}\frac{f_{0}}{1~\rm
kHz}\biggr)^{\!1/2}\biggl(\frac{D_{\rm
L}}{100~\rm Mpc}\biggr)^{\!-1}~\rm.
\label{signal}
\end{eqnarray}
In Figure \ref{fig:LIGO}, we plot the GW strain sensitivity for advanced-LIGO
\cite[aLIGO,][]{Aasi2015} and Einstein Telescope \cite[ET,][]{Punturo2010}. It is clear that the GW
strain of GRB 200219A is below the aLIGO noise curve at $z=0.5$ and $z=0.1$, but it can be detected
by current aLIGO at $z=0.01$. \cite{Abbott2017b} presented a search for GW emission from the
remnant of the binary NS merger GW170817 using data from LIGO and Virgo within a short- and
intermediate-time, but no GW signal from the post-merger remnant is found. From the theoretical
point of view, the GW signal can be detected by current LIGO and Virgo if GRB 200219A is indeed
located at $z=0.01$. Inspired by this point, we want to know whether this possible GW signal can be
found out using data from LIGO and Virgo if GRB 200219A is located at $z=0.01$. By searching
archived data from the LIGO website, we find that there is not any GW signal detected by LIGO in
the interval two hours since the GRB trigger. This is an independent argument to show that the GRB
200219A cannot be located at $z=0.01$. Moreover, the signal may be detected by more sensitivity
instruments at $z=0.01$ and $z=0.1$ in the future, such as ET.

\section{Possible merger-nova emission}
Neutron-rich ejecta can be powered by the merger of an NS binary, and heavier radioactive elements
could be synthesized via r-process \citep{Metzger2017}. \cite{Li1998} first calculated a
near-isotropic signal in the optical/IR band that is powered by radioactive decay (without energy
injection from the central engine). \cite{Yu2013} proposed that an optical/infrared transient can
be powered by merger ejecta for the central magnetar, and this transient is brighter than that of
in \cite{Li1998} due to an additional source of sustained energy injection from the magnetar, they
called it merger-nova \citep{Yu2013,Gao2017}. If this is the case, one interesting question is how
bright this merger-nova is? In this section, following the method of \cite{Yu2013}, we roughly
calculate merger-nova emission at different distances. We adopt the following parameters which are
from the first double NSs merger event (GW170817/GRB 170817A) by fitting the light curve of
AT2017gfo, so that the parameters of merger-nova we used {\bf are} from \cite{Yu2018} and
\cite{Hajela2019}, e.g., ejecta mass $M_{ej}=10^{-2}~M_{\odot}$, velocity $\beta_{ej}=0.1$c,
opacity $\kappa=0.97~\rm cm^{2}~g^{-1}$, medium density $n=0.01~\rm cm^{-3}$, as well as the
spin-down timescale roughly equal to $t_{b,2}$.

\begin{figure}
\centering
\includegraphics [angle=0,scale=0.25] {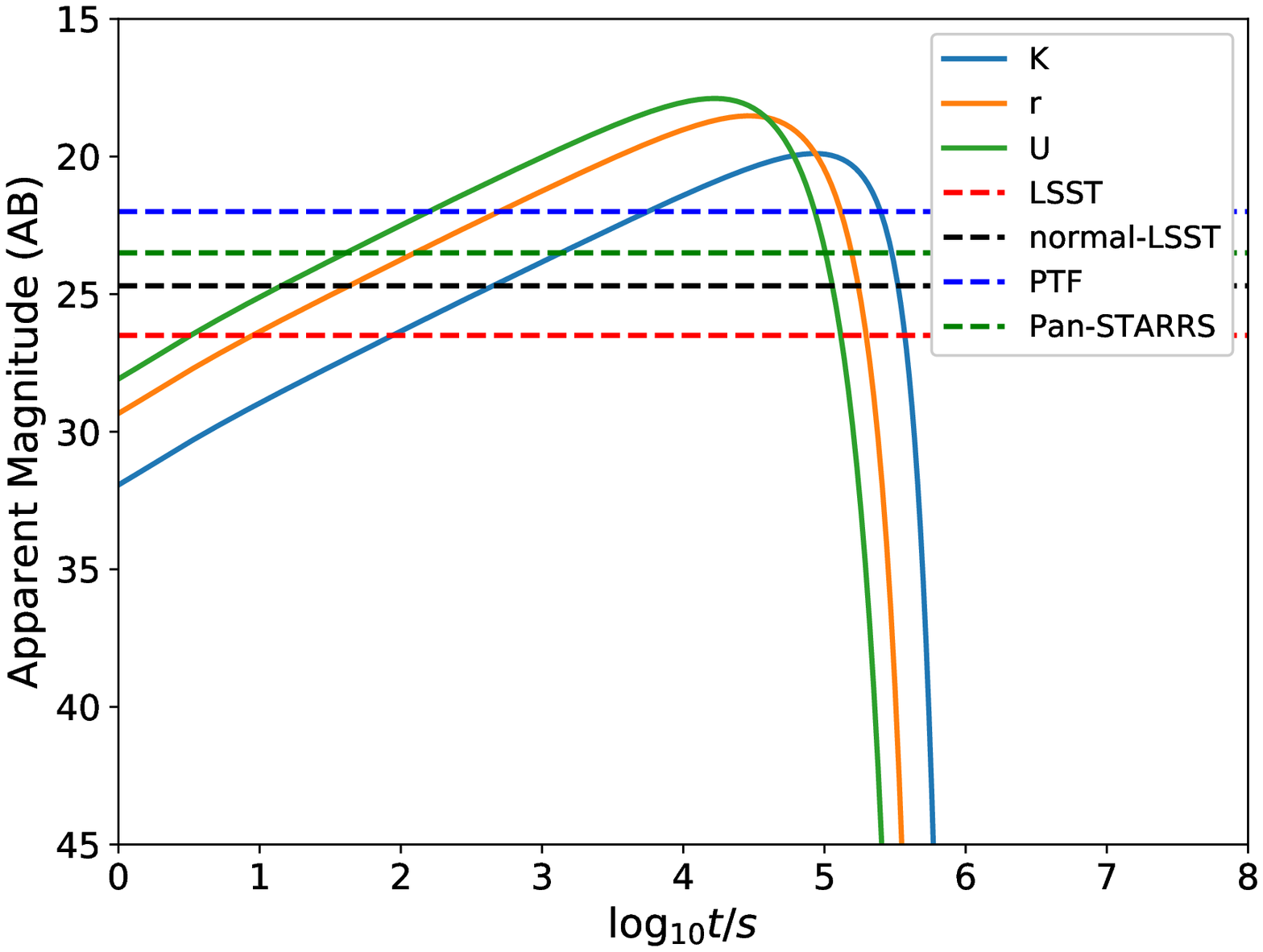}
\includegraphics [angle=0,scale=0.25] {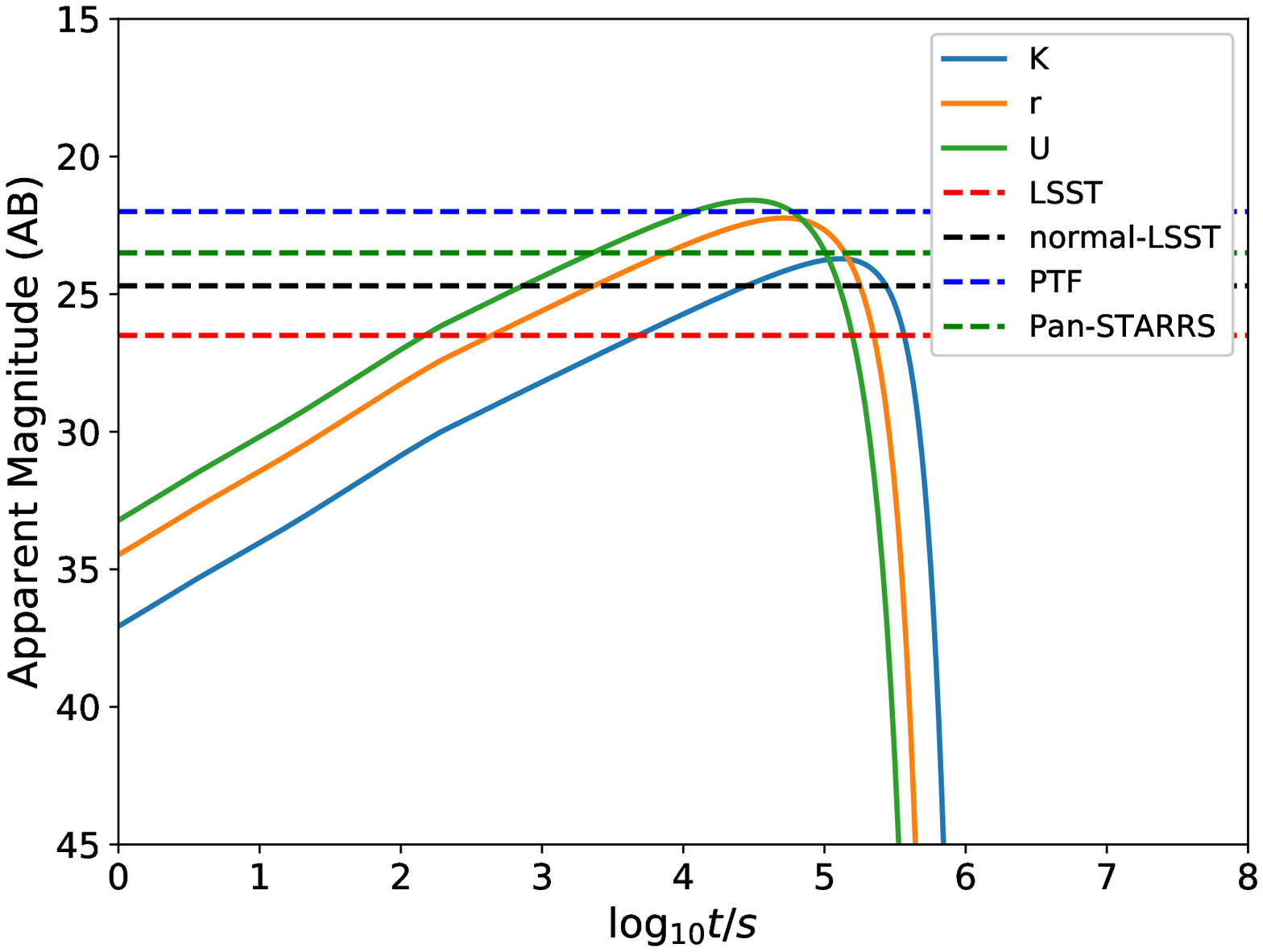}
\includegraphics [angle=0,scale=0.25] {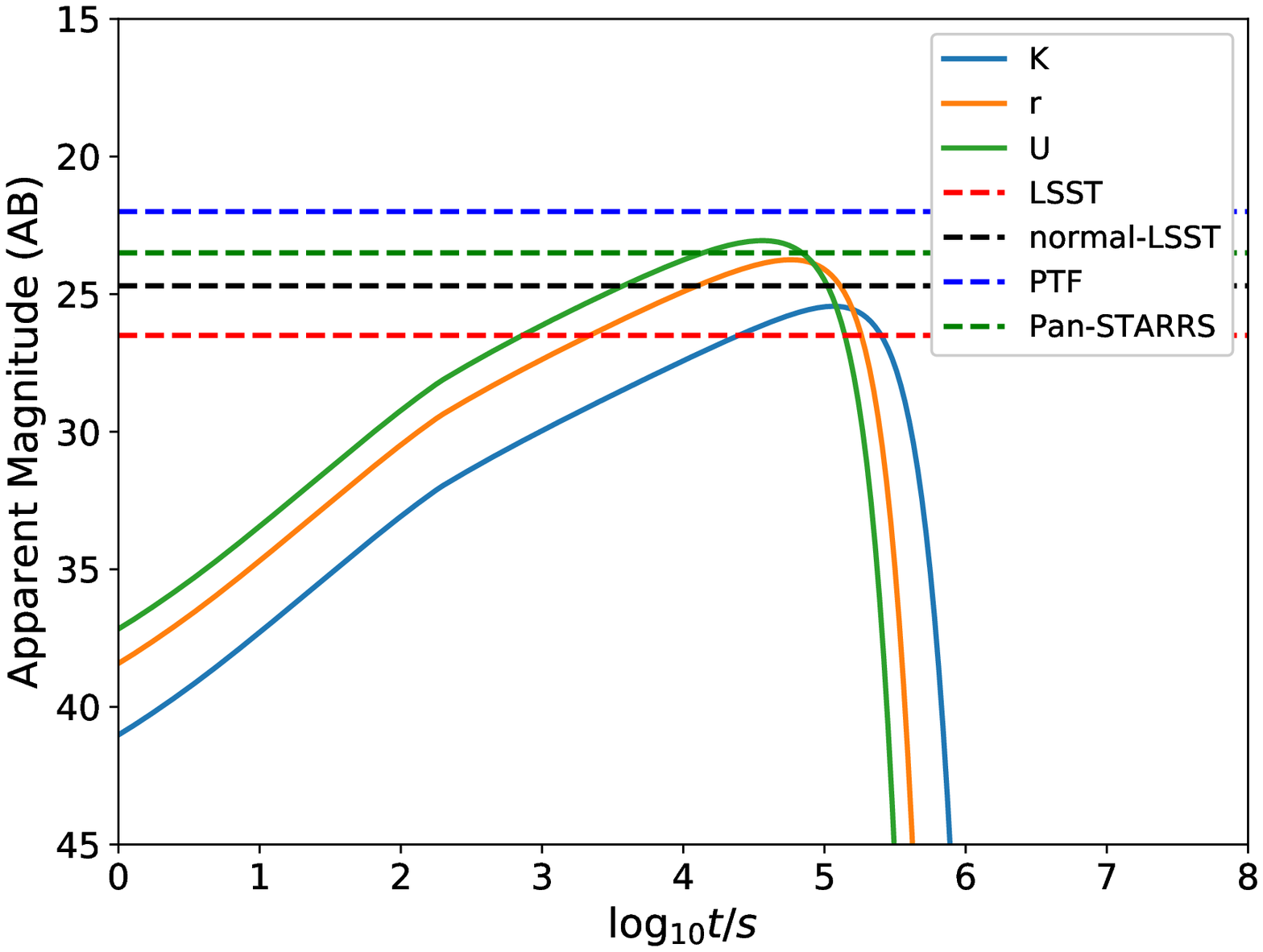}
\caption{Merger-nova light curve of GRB 200219A by only considering the contribution of the magnetar-powered
in K-, r-, and U-band at $z=0.01$ ({\em left}), 0.1 ({\em middle}), and 0.5 ({\em right}).
The horizontal dotted lines are corresponding to upper limit detected of normal-LSST (black), PTF (blue),
Pan-STARRS (green), and LSST (red) surveys, respectively.}
\label{fig:Mergernova}
\end{figure}

Figure \ref{fig:Mergernova} shows the possible merger-nova light curve of GRB 200219A by only
considering the contribution of the magnetar-powered in K-, r-, and U-band at $z=0.01$, 0.1, and
0.5. Moreover, we also overplot the upper limit detected of instruments in Figure
\ref{fig:Mergernova}, e.g., Large Synoptic Survey Telescope(LSST), normal-LSST\footnote{The
expected maximum depth of LSST and normal-LSST are about 26.5 and 24.7 magnitude, respectively. It
means that the LSST should be better to observe more dim image within a longer exposure time
\citep{Metzger2012}.}, PTF, and Pan-STARRS \citep{Jedicke2007,Law2009,Jones2009,Metzger2012}. The
numerical calculation shows that the merger-nova is bright enough to be detected by all instruments
above at $z=0.01$. However, it is a little bit dim at $z=0.1$ and 0.5. More details of
systematically searching and calculations with measured redshift short GRBs are presented in Yuan
et al. (2020).

\section{Discussion}
In the previous works, the observed steep decay (decay slope typically $3\sim5$) after prompt
emission can be interpreted as the curvature effect, which is the delay of propagation of photons
from high latitudes with respect to the line of sight
\citep{Fenimore1996,Kumar2000,Liang2006,Zhang2007}. If this is the case, the predicted temporal
decay index and spectral index of the emission satisfy with a relation,
\begin{eqnarray}
\alpha=2+\beta
\label{Curvature}
\end{eqnarray}
In order to test this possibility, we extract the time-average spectrum of X-ray during the abrupt
drop phase with the power-law model. One has $\beta=\Gamma-1=0.87\pm0.22$, where $\Gamma=1.87\pm0.22$
is the power-law index of spectral fitting. It is easy to check that it is not consistent with
above predicted correlation. Moreover, the plateau emission followed by a $\sim t^{-1}$ decay phase
are also inconsistent with the curvature effect.

Alternatively, the sharp decay of X-ray light curve in the {\em Swift} era was usually interpreted
as jet break, which is geometric effect when the fireball decelerate with the beaming angle
eventually exceeding the physical collimation angle \citep{Liang2008,Racusin2009}. If this sudden
decrease in the flux at $t=200$ seconds after trigger is caused by jet break, the break time is
much shorter than that of {\em Swift} GRBs observed jet break which are as long as $10^4\sim 10^5$
s \citep{Liang2008,Racusin2009}. On the other hand, a normal decay phase with temporal index
$\alpha\sim 1.04$ is observed again after the sharp decay, this segment should not appear if the
sharp decay is interpreted with jet break.

\begin{figure}
\centering
\includegraphics [angle=0,scale=0.6] {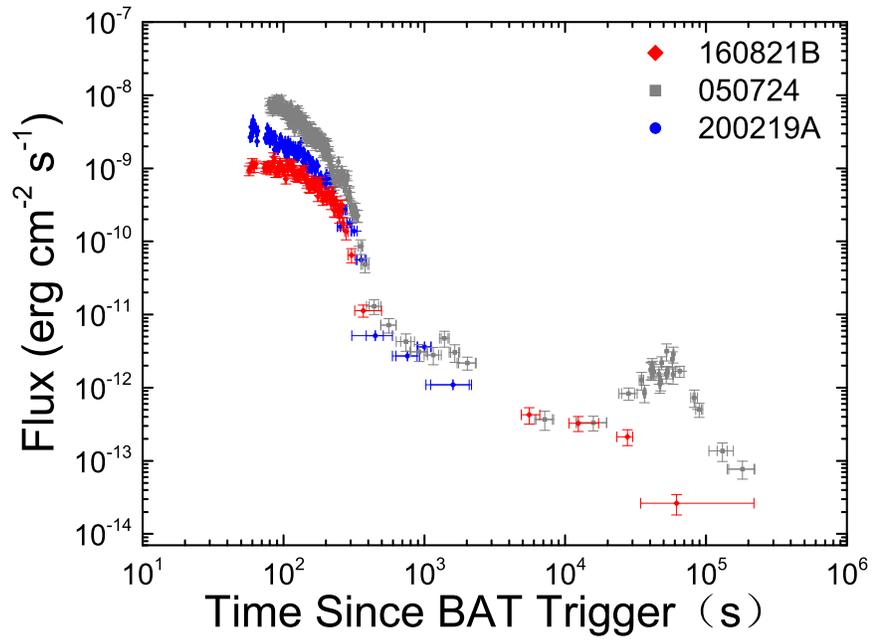}
\caption{Comparison X-ray light curves of GRB 200219A with GRBs 050724 and 160821B in (0.3-10) keV.
The {\bf X-ray} data of GRBs 050724 and 160821B are taken from {\em Swift}/XRT webset: $\rm
https://www.swift.ac.uk/xrt\_curves/allcurves.php$.}
\label{fig:ThreeGRBs}
\end{figure}

By systematically searching all short GRBs observed with {\em Swift}, we find that another two
short GRBs with redshift measured are also consistent with the magnetar central engine, but the
physical process is different from GRB 200219A. (1) One is GRB 050724 with EE at redshift $z=0.258$
\citep{Barthelmy2005}, its early X-ray light curve presents a plateau emission followed by a $\sim
t^{-2}$ decay phase, then continue to an abrupt drop segment ($\sim t^{-8}$). It is natural to
explain by invoking supramassive magnetar from the merger of an NS binary. The magnetar spins-down
losing its rotation energy mostly via the EM channel and then collapse into a black hole after
surviving hundreds of seconds. (2) Another one is nearby GRB 160821B with $z=0.16$
\citep{Levan2016}; its early X-ray light curve shows a plateau emission followed by an abrupt drop
decay ($\sim t^{-4.5}$). However, there is no signature revealing that whether the collapse is
caused by EM or GW dominated radiation, more details also see \cite{Lyu2017}. A comparison of the
X-ray light curves of those three short GRBs are shown in Figure \ref{fig:ThreeGRBs}. These results
suggest that at least a supramassive NS/magnetar can survive in the central engine of some short
GRBs, and it spins down via losing its rotation energy due to either GW-dominated radiation or
EM-dominated radiation. Moreover, \cite{Sarin2019} found that the millisecond magnetar model is
favoured over the fireball model for two short GRBs by analyzing its X-ray data, but it is
dependent on the unknown equation of state and non-rotating neutron star mass. The more robust
evidence of this hypothesis is to catch such weak GW signal from supramassive magnetar by a-LIGO
and Virgo in the future.

\section{Conclusions }
GRB 200219A is a short GRB with duration less than 1 s, observed by both {\em Swift} and {\em
Fermi}. The ¡°extended emission¡± component lasting $\sim 90$ s after the initial hard spike is
identified by the {\em Swift}/BAT, but it is not significant in the {\em Fermi}/GBM temporal
analysis. We presented a broadband analysis of its prompt and afterglow emission and found that the
peak energy of its spectrum is as high as $1387^{+232}_{-134}$ keV, which is harder than most short
GRBs observed by {\em Fermi}/GBM. More interestingly, together with the EE component and early
X-ray data, a plateau emission was followed by a $\sim t^{-1}$ segment, then with an extremely
steep decay. This early temporal feature is very difficult to explain with the standard
internal/external shock model of a black hole central engine, but could be consistent with the
prediction of a magnetar central engine from the merger of an NS binary. We explain the plateau
emission followed by a $\sim t^{-1}$ decay phase from spinning down of millisecond magnetar, which
loses its rotation energy via GW quadrupole radiation. Then, the magnetar collapsing into the black
hole before switching to the EM-dominated is corresponding to abrupt drop decay.

However, a fly in the ointment is no redshift measured of this case, so we have to attempt the
pseudo redshift to reveal its physical properties. Several numerical calculations are summarized as
following:
\begin{itemize}
 \item By assuming that the pseudo redshift $z=0.01$ which is the approximate luminosity
     distance as GW170817/GRB 170817A event. The requirements of physical parameters of
     magnetar are not reasonable, especially, $B_p$ is as high as $10^{17}$ G. The signal of GW
     radiation of central magnetar at this distance is high enough and can be detected by
     current a-LIGO and ET in the future. Moreover, at a later time, the peak luminosity of
     possible merger-nova is also above the upper limit detected of several instruments.
\item If the pseudo redshift $z=0.1$, that is close to the upper limit of GW signal detected by
    current LIGO/Virgo. The physical parameters of magnetar seem to be in a reasonable range.
    The signal of GW at this distance is below the noise curve of current a-LIGO, but is
    expected to detect by ET in the future. The merger-nova signal is also potential to be
    detected by survey telescopes in the future.
\item If we adopt the pseudo redshift $z=0.5$, which is the central value of the redshift
    distribution for all short GRBs with $z$ measurements, this requires a rapidly rotating
    magnetar with a spin period $\sim 1$ ms, and a surface magnetic field in a reasonable
    range. The GW signal cannot be detected by a-LIGO or ET. The merger-nova signal at this
    distance can be comparable with the limits from some optical survey telescopes.

 \end{itemize}

\begin{acknowledgements}
We acknowledge the use of public data from the {\em Swift} and {\em Fermi} data archive, and the UK
{\em Swift} Science Data Center. This work is supported by the National Natural Science Foundation
of China (grant Nos.11922301, 11851304, 11533003, and 11833003), the Guangxi Science Foundation
(Grant Nos. 2017GXNSFFA198008, 2018GXNSFGA281007, and AD17129006). The One-Hundred-Talents Program
of Guangxi colleges, Bagui Young Scholars Program (LHJ), and special funding for Guangxi
distinguished professors (Bagui Yingcai \& Bagui Xuezhe). BBZ acknowledges support from a national
program for young scholars in China, Program for Innovative Talents and Entrepreneur in Jiangsu,
and a National Key Research and Development Programs of of China (2018YFA0404204).

\end{acknowledgements}

%********************************References*****************************************************

%*************************************************************************************
%********************************Table1*****************************************************
\clearpage
\begin{center}
\begin{deluxetable}{cccccccccccccc}
%\rotate
\tablewidth{0pt} \tabletypesize{\footnotesize}
%\tabletypesize{\tiny}
\tablecaption{Curved Power-law spectral fit parameters for the {\em Fermi}/GBM $T_{90}$ data of GRB
200219A} \tablenum{1}
\tablehead{Time Interval& & & CPL & &\\
\hline \colhead{$t_s$}& \colhead{$t_e$}& \colhead{$\Gamma_{\rm ph}$} & \colhead{$E_{\rm p}$} &
\colhead{$logNorm$}&\colhead{PGSTAT/dof}} \startdata
-0.02 & 0.5 & $-0.65\pm0.07$ & $1387_{-134}^{+232}$ & $-0.19_{-0.15}^{+0.13}$ & 278/351\\
\enddata
\end{deluxetable}
\end{center}

%********************************Table2*****************************************************

\begin{center}
\begin{deluxetable}{cccccccccccccc}
%\rotate
\tablewidth{0pt} \tabletypesize{\footnotesize}
%\tabletypesize{\tiny}
\tablecaption{The derived parameters of magnetar for different redshift} \tablenum{2}

\tablehead{ \colhead{Redshift}& \colhead{$L_{em,0}$}& \colhead{$B_p$} & \colhead{$P_0$} &
\colhead{$h_c$\tablenotemark{a}}\\
\colhead{($z$)}& \colhead{($\rm erg~s^{-1}$)}& \colhead{(G)} & \colhead{($10^{-3}$ s)} &
\colhead{}}
\startdata
0.01 &$(1.09\pm 0.24)\times 10^{46}$ &$<4.6\times 10^{17}$  &$<118$  &$~1.96\times 10^{-23}$ \\
0.1  &$(1.30\pm 0.31)\times 10^{48}$ &$<4.3\times 10^{16}$  &$<11.2$ &$~1.79\times 10^{-24}$ \\
0.5  &$(4.92\pm 1.13)\times 10^{49}$ &$<9.4\times 10^{15}$  &$<2.1$  &$~2.91\times 10^{-25}$ \\
\enddata
\tablenotetext{a}{The GW strain of magnetar for initial $f_0=1000$ Hz.}
\end{deluxetable}
\end{center}

%********************************Figure1*****************************************************
%%----------------------------------------------------------------------------
%%----------------------------------------------------------------------------

%********************************Figure2*****************************************************
%\begin{figure}
%\centering
%\includegraphics [angle=-90,scale=0.3] {SEE.eps}
%\includegraphics [angle=-90,scale=0.3] {SBG.eps}
%\caption{The spectra of EE component (left) and background (right) with the power-law model fits.}
%\label{fig:SpecBAT}
%\end{figure}
%----------------------------------------------------------------------------
%----------------------------------------------------------------------------

%----------------------------------------------------------------------------
%********************************Figure3*****************************************************
%%----------------------------------------------------------------------------
%%----------------------------------------------------------------------------
%********************************Figure4*****************************************************
%%----------------------------------------------------------------------------
%********************************Figure5*****************************************************
\begin{figure}
\centering
\includegraphics [angle=0,scale=0.35] {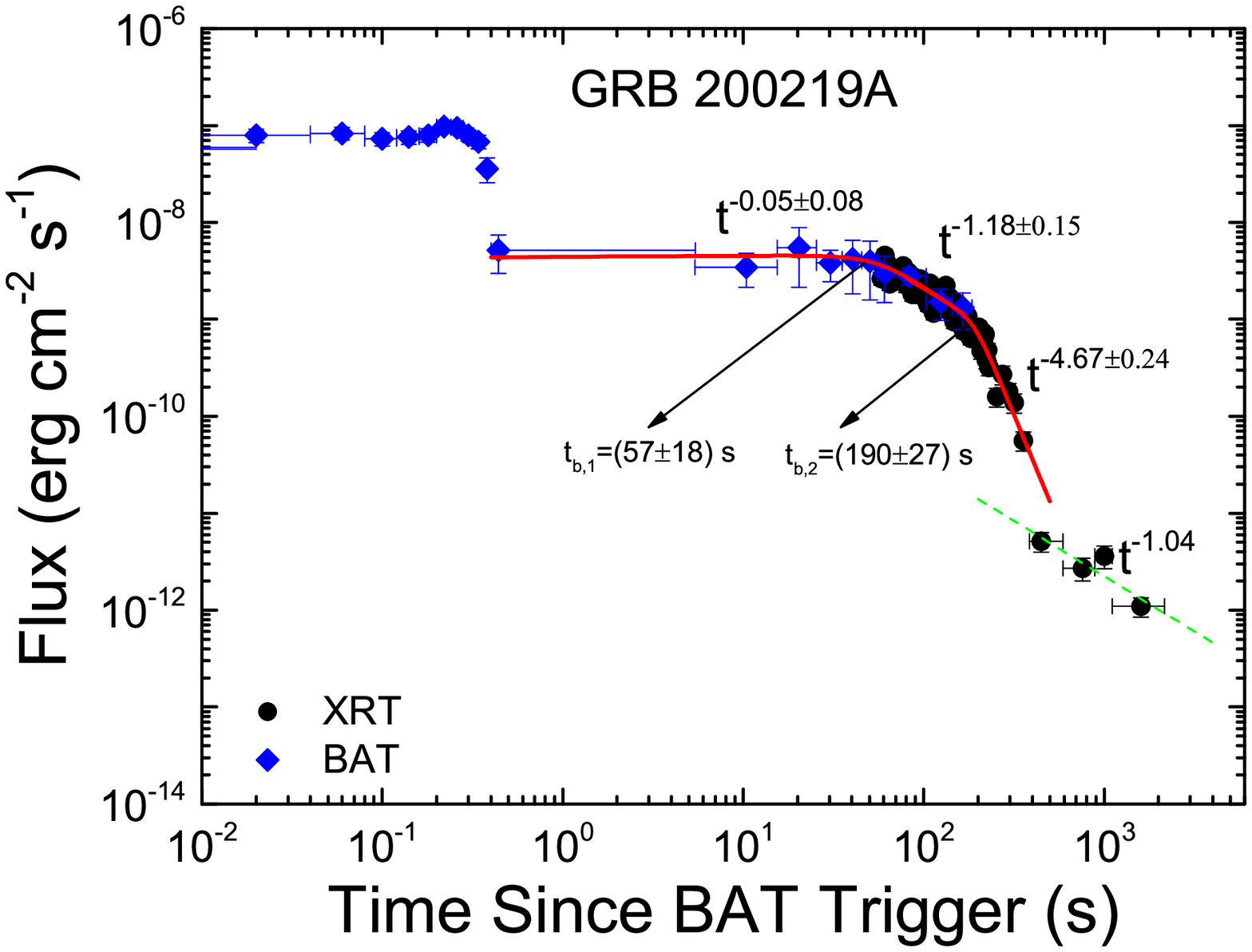}
\includegraphics [angle=0,scale=0.35] {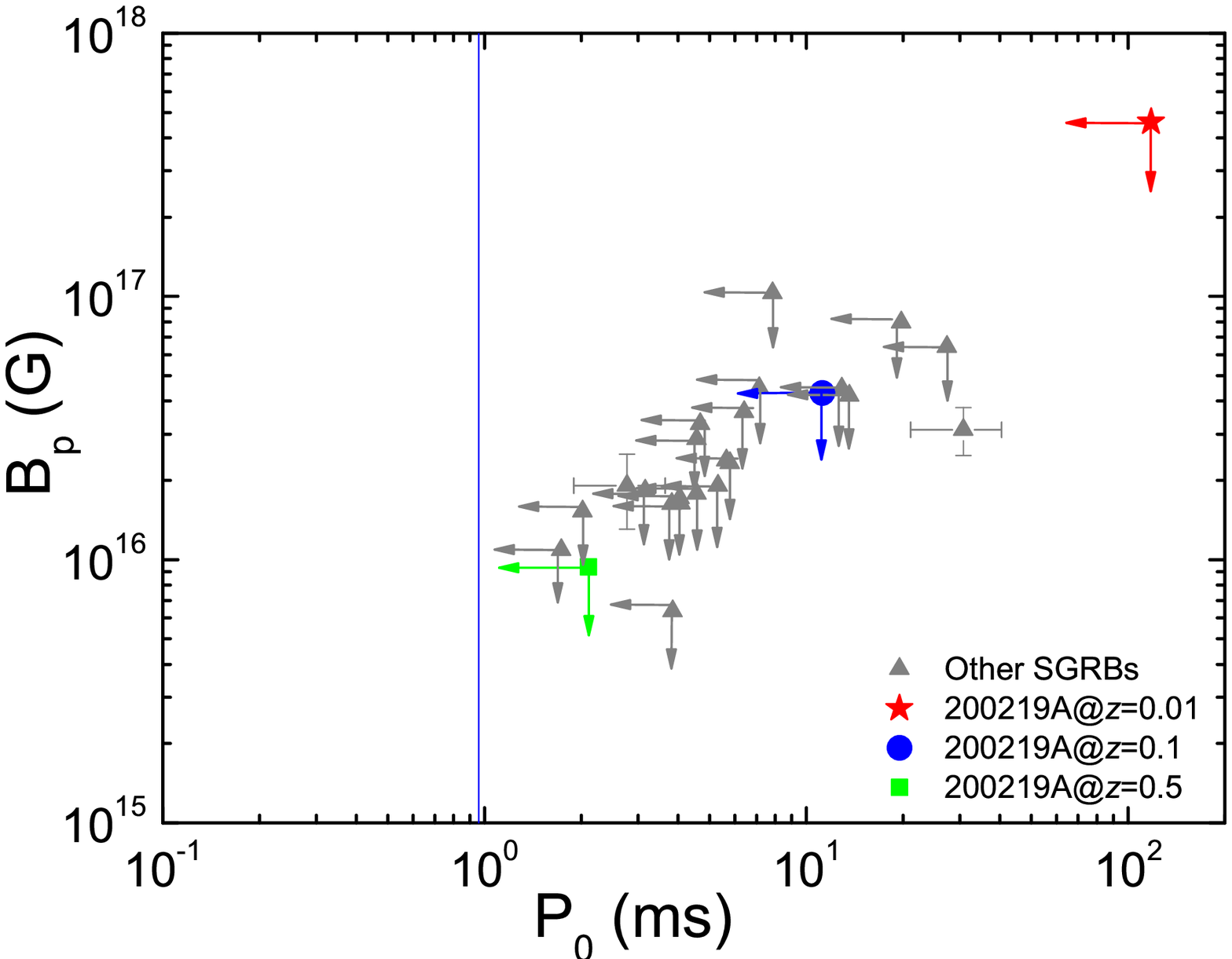}
\caption{{\em Left}: X-ray light curve in (0.3-10) keV and empirical
fit with a smoothed triple power law model.
{\em right}: Inferred magnetar parameters ($P_0$ vs. $B_p$) of GRB 200219A with $z=0.01$ (red star),
0.1 (blue dot), and 0.5 (green square) compared with other short GRBs (gray triangle).
The derived magnetar parameters of other short GRBs are taken from L\"{u} et al. (2015).
Vertical solid line is the break-up spin period limit of neutron star.}
\label{fig:XRT}
\end{figure}
%%----------------------------------------------------------------------------
%********************************Figure6*****************************************************
%%----------------------------------------------------------------------------
%********************************Figure7*****************************************************
%%----------------------------------------------------------------------------
%********************************Figure8*****************************************************
%%----------------------------------------------------------------------------

\end{document}